# Intrinsic skyrmions in monolayer Janus magnets

Jiaren Yuan,[1,†] Yumeng Yang,[2,†] Yongqing Cai,[3] Yihong Wu,[2] Yuanping Chen,[1] Xiaohong Yan,[1,*] Guiyin Xu,[4,*] and Lei Shen[5,*]

[1]*College of Science and School of Material Science and Engineering, Jiangsu University, Zhenjiang 212013, China*

[2]*Department of Electrical and Computer Engineering, National University of Singapore, 4 Engineering Drive 3, Singapore 117583*

[3]*Joint Key Laboratory of the Ministry of Education, Institute of Applied Physics and Materials Engineering, University of Macau, Macau 999078, China*

[4]*Department of Nuclear Science and Engineering, Massachusetts Institute of Technology, Cambridge, Massachusetts 02139, USA*

[5]*Department of Mechanical Engineering and Engineering Science, National University of Singapore, 9 Engineering Drive 1, Singapore 117542*

Skyrmions are localized solitonic spin textures with protected topology, which are promising as information carriers in ultra-dense and energy-efficient logic and memory devices. Recently, magnetic skyrmions have been observed in magnetic thin films, and are stabilized by the *extrinsic* interfacial Dzyaloshinskii-Moriya interaction (DMI) and/or external magnetic fields. The specific effects in magnetic monolayer materials have not been thoroughly studied. Here, we investigate the *intrinsic* magnetic skyrmions in a family of monolayer Janus van der Waals magnets, MnSTe, MnSeTe, VSeTe, and MnSSe, by the first-principles calculations combined with the micromagnetic simulations. The monolayer Janus MnSTe, MnSeTe, and VSeTe with out-of-plane geometric asymmetry and strong spin-orbit coupling (SOC) have a large intrinsic DMI, which could stabilize a sub-50 nm intrinsic skyrmions in monolayer MnSTe and MnSeTe at zero magnetic field. While monolayer VSeTe with in-plane easy axis forms magnetic domain rather than skyrmions. Moreover, the size and shape of skyrmions can be tuned by an external magnetic field. Therefore, our work motivates a new vista for seeking intrinsic skyrmions in atomic-scale magnets.

# I. INTRODUCTION

Skyrmions have rapid development in recent years [1-3]. The generation and motion of skyrmions are induced and driven by the injected spin current [4-6] through the spin Hall effect [7,8] or Rashba-Edelstein effect [9]. These phenomena enable all-electric controlled skyrmion systems to effectively manipulate the spin, providing potential applications for new memories and logic devices with high density and high performance [10-12]. Skyrmions are originally observed in magnetic cubic B20 bulk crystals with no centrosymmetry, such as MnGe[13] and MnSi [14]. Subsequently, atomic-scale skyrmions are observed in ultrathin epitaxial magnetic Fe/Ir(111) bilayers [15] at low temperature due to a significant interfacial Dzyaloshinskii-Moriya interaction (DMI). The interfacial DMI plays a crucial role in the creation, stabilization and manipulation of chiral domains and skyrmions at the interface between $3d$ ferromagnetic and heavy $5d$ nonmagnetic metals [6]. These interfaces carry a sizable DMI due to the structural symmetry-breaking and the strong SOC [16]. Recently, room-temperature skyrmions have been stabilized by stacking the bilayers into multilayers, leveraging more interfacial DMI [17,18]. However, the interfacial DMI is affected by the interfacial defects and atomic stacking order [18]. Therefore, it is important and urgent to search monolayer two-dimensional (2D) magnetic materials with intrinsic DMI to avoid the interfacial disadvantages above.

Recently, a large amount of 2D intrinsic van der Waals (vdW) magnets have been synthesized, such as $MnSe_2$ [19], $VSe_2$ [20], $Fe_3GeTe_2$ [21] and $CrI_3$ [22]. It provides opportunities to find 2D hosts with intrinsic DMI and skyrmions. However, all these 2D magnets have no broken inversion symmetry, which is essential for the creation of DMI and skyrmions [23,24]. There are several ways to break the inversion symmetry in 2D systems, such as the electric field, Moiré rotation, doping, and heterostructure. For example, Liu *et al*. applied an electric field on $CrI_3$ to break the symmetric potential field, thereby obtaining DMI [25]and skyrmions [26]. Tong *et al*. reported skyrmions in Moiré 2D vdW ferromagnetic materials [27]. 2D magnets without the inversion symmetry, and with the intrinsic DMI and skyrmions are more desirable for practical applications. The reported Janus monolayer transition metal chalcogenides (TMDs) with inversion asymmetry provided a strategy for designing intrinsic DMI and skyrmions [28,29]. However, the synthesized 2D Janus MoSSe and WSSe are non-magnetic [30]. Theoretically, several Janus TMDs are predicted to be energetically stable with intrinsic ferromagnetism [31,32]. Therefore, it is meaningful to examine whether large intrinsic DMI and stable skyrmions can be obtained in such magnetic Janus TMDs with broken inversion symmetry and strong SOC.

Motivated by the reported monolayer Janus TMDs and room temperature magnets $MnSe_2$ and $VSe_2$, we investigated the magnetic properties, including the magnetic exchange interaction, SOC and DMI, of four Janus vdW magnets (MnSTe, MnSeTe, VSeTe, and MnSSe) by first-principles calculations, and studied the generation and regulation of skyrmion states using micromagnetic simulations. A significant intrinsic DMI was found in three Janus MnSTe, MnSeTe, and VSeTe due to the lack of inversion symmetry and the presence of large SOC. The results of micromagnetic simulation showed that skyrmions could be stabilized in MnSTe and MnSeTe in the absence of an external magnetic field because of their smaller $K_u$ compared with $1.25 \frac{D^2}{A}$, where $D$ was the value of micromagnetic DMI constant, $A$ was the exchange stiffness, and $K_u$ was the magnetic anisotropy constant. VSeTe only showed the magnetic domains with no skyrmions because of its large $K_u$. It is found that the size of individual skyrmions in MnSTe and MnSeTe is 40 and 41 nm, respectively, which can be further shrunk into sub-10 nm *via* an external magnetic field. Therefore, our study provides an alternative for the generation of skyrmions in 2D vdW magnets, guiding the direction of realizing the high-density spin-orbitronic devices.

## II. METHODOLOGY

All the first-principles calculations were performed by density functional theory (DFT) based Vienna *ab initio* Simulation Package (VASP) [33]. We employ a plane-wave basis set [34] with a cutoff energy of 450 eV. Perdew-Burke-Ernzerhof (PBE) within the generalized gradient approximation (GGA) is chosen for the exchange-correlation functional [35]. The DFT+U method [36] is carried out to deal with electron correlations for transition metal Mn atom with $U_{eff}$=3.9 eV, respectively. The monolayer Janus magnetic TMDs are simulated by a slab model with a 15 Å vacuum layer. The electron energy is computed using a 21×21×1 $k$-point mesh for slab geometries. 6×24×1 $k$-points are utilized for the 4×1 supercells. The convergence criterion is set to be $10^{-6}$ eV. The structures are fully relaxed until the force for all atoms is less than $10^{-3}$ eV/Å. The Tersoff-Hamann method is implemented to simulate the scanning tunneling microscopy (STM) with constant height mode [37], and the STM tip is located above the surface with a distance of 3 Å.

We performed micromagnetic simulations using the MuMax3 software package [38]. The energy of ferromagnetic systems is derived as the sum of magnetic exchange energy ($E_{ex}$), antisymmetric exchange energy ($E_{DMI}$), magnetic anisotropy energy ($E_{ani}$), Zeeman energy ($E_{Zeeman}$) and magnetostatic energy ($E_{demag}$), $E_{tot} = E_{ex} + E_{DMI} + E_{ani} + E_{zeeman} + E_{demag}$. The monolayer Janus magnetic TMDs are modeled with a mesh containing 1024 × 1024 × 1 cells. The cell size is 1×1× 0.4 nm. The length of $z$ direction is taken as the effective thickness of monolayer Janus structure. The parameters of the magnetic anisotropic constant ($K_u$), the DMI values and exchange interaction $A_{ex}$ are derived from the DFT calculations. The simulations are initialized with a single Néel skyrmion embedded in a ferromagnetic background. The ferromagnetic spin arrangement is set as +z and the spin arrangement of skyrmion core is set as -z. For the given initial condition, the equilibrium magnetization configuration with minimum total energy is sought by successive iterations. The final magnetization state is reached until its total energy is converged.

### III. RESULTS AND DISCUSSION

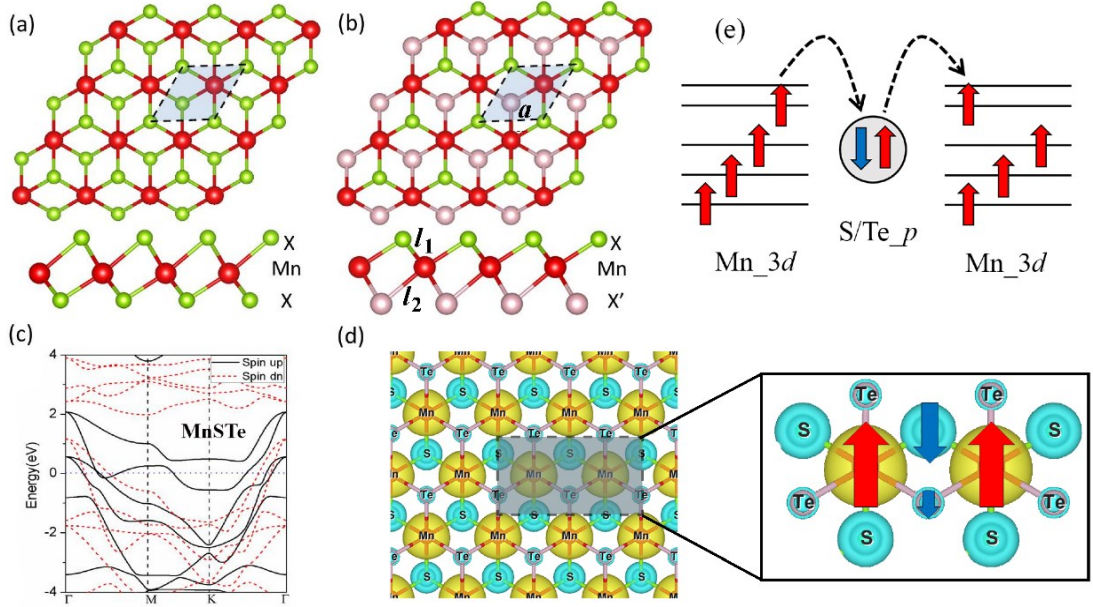

**FIG. 1.** Atomic structures of pristine and Janus TMDs: (a) Top and side view of pristine TMDs, (b) Top and side view of Janus TMDs. Green X and pink X' balls represent different chalcogen atoms (S, Se and Te), while the red ball denotes the Mn atom. (c) The spin-polarized band structure of MnSTe. The spin up and spin down bands are indicated by black solid lines and red dashed lines, respectively. The Fermi level is set to zero and is represented by a blue dotted line. (d) The spin-polarized charge density of MnSTe. The spin up and spin down charge are indicated by yellow and blue isosurface, respectively. The collinear spin orientation on Mn and S/Te is shown in the enlarged Fig. (e) The schematic diagram of the Zener double-exchange mechanism for MnSTe.

The geometric structures of monolayer pristine (a) and Janus (b) TMDs are shown in **Fig. 1**. Janus structures have broken out-of-plane structural symmetry with different bond lengths, $l_1$ and $l_2$. The optimized structural parameters are summarized in **Table 1**. These Janus structures are stable as discussed in **Fig. 5**. In this work, we keep the same layered arrangement, *i.e.*, the ingredient of the top atomic layer is the light element X, while the heavy element X' is on the bottom. In all of this work except the particular description, we only take MnSTe as an example because all four monolayer Janus materials have the same geometric structure and symmetry ($C_{3v}$).

There are only three atoms located in the octahedral unit cell, and each Mn atom is bonded to six chalcogen atoms. In the perfect octahedral crystal field, the 3$d$ orbitals of the Mn atom split into 2-fold $e_g$ and 3-fold degenerate $t_{2g}$ orbitals. In the Janus MnSTe, which is lack of inversion symmetry, the different chalcogen atoms (S and Te) induce the non-degenerated $d$ states of Mn and a high-spin state [see **Fig. 1(c)** and **1(e)**]. The high spin-polarized states induce large magnetic moments in Janus MnSTe [see in **Table 1**]. The total magnetic moments of the monolayer MnSTe are 3.51 $\mu_B$ per unit cell and the on-site moments of the Mn atom are 4.12 $\mu_B$. The decreased total moment is due to the negative sign of the magnetic moments of the linked chalcogen atoms as shown by the spin density ($\rho_\uparrow$-$\rho_\downarrow$) [**Fig. 1(d)**]. The spin-up density for Janus MnSTe is largely distributed on the Mn atoms and a few spin-down densities on the S/Te atoms due to the coupling with the Mn atoms. The spin directions of the chalcogen atoms are opposite to those of Mn atoms, thus the Mn atoms maintain an antiferromagnetic coupling through their surrounding chalcogen atoms. This is a typical Zener double-exchange interaction in conducting Mn-based magnets as the schematic of this exchange mechanism shown in **Fig. 1(e)**.

**Table 1.** Structural and magnetic parameters of Janus TMDs. the lattice constant, $a$, the bond lengths between transition metal and chalcogen atom, $l_1$ and $l_2$, the magnetic moment of the transition metal atom, $M_{TM}$, the total magnetic moment, $M_{tot}$, the in-plane DMI, $d_{//}$, micromagnetic DMI coefficient, $D$, isotropic exchange, $J$, exchange stiffness, $A$, magnetic anisotropic constant, $K_u$.

|  | $a$ (Å) | $l_1$ (Å) | $l_2$ (Å) | $M_{TM}$ ($\mu_B$) | $M_{tot}$ ($\mu_B$) | $d_{//}$ (meV) | $D$ (mJ/m²) | $J$ (meV) | $A$ (pJ/m) | $|K_u|$ (J/m³) | $1.25*|D^2/A$ (J/m³) |
|---|---|---|---|---|---|---|---|---|---|---|---|
| MnSTe | 3.61 | 2.41 | 2.85 | 4.12 | 3.51 | 5.58 | 9.1 | 6.4 | 11.2 | 6.0×10⁶ | 9.2×10⁶ |
| MnSeTe | 3.71 | 2.55 | 2.81 | 4.11 | 3.52 | 4.34 | 6.7 | 12.9 | 13.3 | 4.1×10⁶ | 4.2×10⁶ |
| VSeTe | 3.60 | 2.52 | 2.79 | 1.96 | 1.55 | 1.25 | 2.0 | 2.2 | 3.9 | 4.4×10⁶ | 1.3×10⁶ |
| MnSSe | 3.56 | 2.41 | 2.54 | 3.87 | 3.18 | -0.04 | -0.01 | 25.1 | 12.6 | 5.6×10⁵ | 0.0 |

We consider three energy terms in the spin Hamiltonian for the Janus magnetic systems as,

$$\widehat{H}_{spin} = -\sum_{ij} J_{ij} \mathbf{S}_i \cdot \mathbf{S}_j - \sum_i K_{zz} S_{iz}^2 - \sum_{ij} \mathbf{d}_{ij} \cdot (\mathbf{S}_i \times \mathbf{S}_j), \qquad (1)$$

where $\mathbf{S}_i$ ($\mathbf{S}_j$) is the spin vector on the $i^{th}$ ($j^{th}$) site, $J_{ij}$ is the isotropic exchange, $K_{zz}$ is the easy-axis single ion anisotropy, $\mathbf{d}_{ij}$ is the anisotropic antisymmetric exchange, respectively. The terms of the external magnetic field and magnetostatic dipolar interaction are not included. It is noted that we only consider the first nearest neighboring pair of Mn atoms. These coefficients can be extracted from the total energy difference method [16].

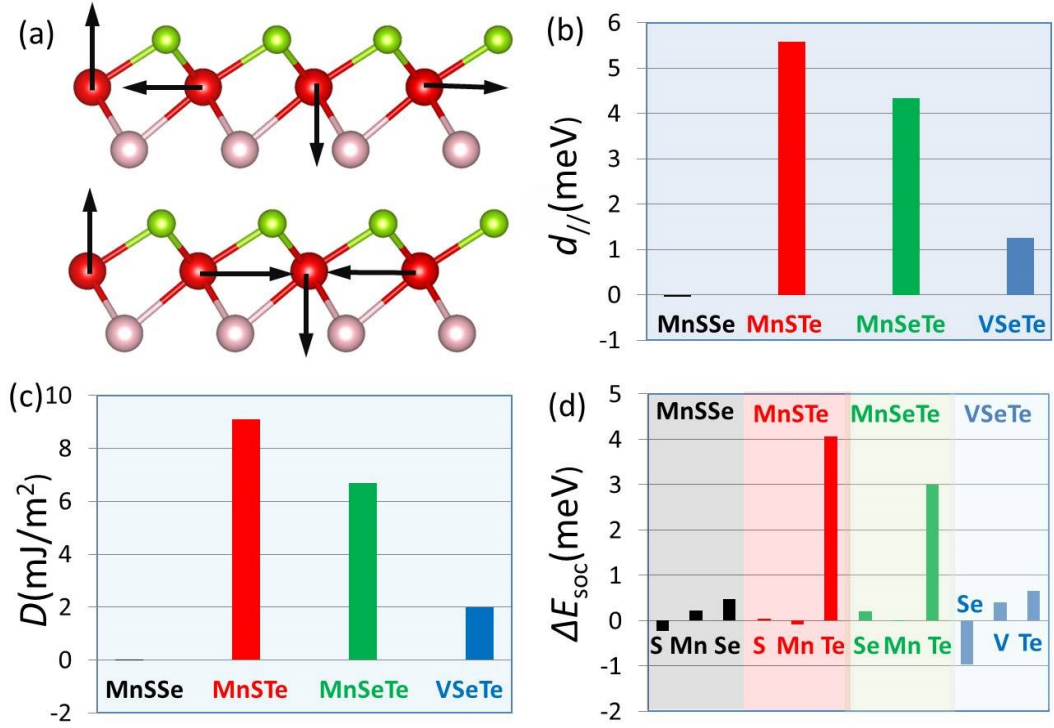

**FIG. 2.** (a) Left-hand and right-hand spin-spiral configurations used to calculate the DMI and $\Delta E_{SOC}$ of Janus TMDs. (b) The calculated microscopic DMI $d_{//}$, (c) the micromagnetic DMI $D$, and (d) atomic resolved localization of the associated SOC energy $\Delta E_{SOC}$, defined as the difference of SOC energies between left-hand and right-hand spin-spiral configurations.

The noncollinear magnetic order of the Janus magnetic monolayer, left-hand and right-hand spin-spiral configurations, is shown in **Fig. 2(a)**, which is used to calculate the microscopic ($d$) and micromagnetic ($D$) DMI constants. According to the Moriya' the symmetry selection rules, the Dzyaloshinskii-Moriya vector $\boldsymbol{d}_{ij} = d_{//}(\boldsymbol{z} \times \boldsymbol{u}_{ij}) + d_{\perp}\boldsymbol{z}$, where $\boldsymbol{z}$ is unit vector along the z-axis direction and $\boldsymbol{u}_{ij}$ is the unit vector from site $i$ to site $j$, respectively [16]. The contribution of $d_{\perp}$ to the Néel skyrmions is negligible, hence we only evaluate the values of $d_{//}$. The DMI for one Mn atom from the nearest six magnetic Mn atoms is equal to $\pm\frac{3}{2}d_{//}$ (the sign $\pm$ corresponds to left-hand and right-hand spin-spiral configurations, respectively). For a 4×1 supercell with four Mn atoms, the following relationship $E_L - E_R = 4 \times [\frac{3}{2}d_{//} - (-\frac{3}{2}d_{//})]$ is obtained. Hence, the in-plane component of the microscopic DMI is $d_{//} = \frac{(E_L - E_R)}{12}$. Subsequently, $\boldsymbol{d}_{ij}$ is converted into the micromagnetic DMI, $D$, with the expression: $D =$

$\frac{\sqrt{3}(d_{//})}{at}$, where $a$ is the lattice parameter and $t$ is the effective thickness [6,16].

The calculated microscopic $d_{//}$ and micromagnetic $D$ of DMI for the Janus TMDs are shown in **Fig. 2(b)** and **2(c)**. There are significant values of $d_{//}$ and D for the Janus structures because the symmetry is broken by the difference out-of-plane elements. The values of $d_{//}$ for the Janus MnSSe, MnSTe, MnSeTe, and VSeTe are -0.04, 5.58, 4.34, and 1.25 meV, respectively. The $d_{//}$ of monolayer MnSSe is a tiny negative value. The negligible DMI in MnSSe is due to its quite weak SOC. The extracted $D$ for the Janus MnSTe, MnSeTe, and VSeTe are 9.1, 6.7, and 2.0 mJ/m$^2$, respectively. The $D$ for monolayer MnSTe and MnSeTe is compared to that of Ir/Fe/Co/Pt [18] and graphene/Co interface [39]. It is instructive to evaluate the associated SOC energy source to elucidate the origin of DMI, which is defined as the difference of SOC energies extracted from opposite chiralities in monolayer Janus structure. The atomic resolved localization of the associated SOC energy $\Delta E_{SOC}$ is depicted in **Fig. 2(d)**. Importantly, it is clearly shown that the large DMI for MnSTe, MnSeTe and VSeTe mainly originates from Te atom rather than from the TM atom and other chalcogen atoms.

Other important parameters, such as the isotropic exchange $J$, magnetic anisotropic constant $K_u$ and the exchange stiffness $A$, are calculated and summarized in **Table 1**. The magnetic anisotropy energy (MAE) is calculated as the energy difference between in-plane and out-of-plane magnetization. MnSTe and MnSeTe behave perpendicular magnetic anisotropy with the easy axis along the out-plane axis, while VSeTe has an in-plane easy axis. The magnetic anisotropic constant $K_u$ of MnSTe, MnSeTe, and VSeTe are 6.0×10$^6$, 3.8×10$^6$, and -4.4×10$^6$ J/m$^3$, respectively. The exchange stiffness $A$ is an important parameter which controls the magnetization reversal of the magnetic materials. Based on the relationship between the micromagnetic free energy and the exchange energy, the exchange stiffness $A$ is defined as $A = \frac{U}{V(|\nabla M|^2)}$ [18], where $V$ is the volume of the unit cell, **M** is the unit magnetic vector. $U$ is the exchange energy defined as the difference between the average total energy of noncollinear left-hand and right-hand spin spirals, and the total energy of the collinear spin configuration. The exchange stiffness $A$ of the Janus MnSTe and MnSeTe are 11.2 and 13.3 pJ/m, respectively. These values are very close to the exchange stiffness of magnetic thin film heterojunction in the experiment [18, 40].

Next, we further discuss the physical meanings of the magnetic parameters as listed in **Table 1**. First of all, the positive values of *J* imply ferromagnetism for all four Janus monolayers. In the case of MnSSe, it has a smaller *D* and $K_u$ than the other Te-based compounds due to weaker SOC strength. More importantly, the other Mn-based Janus monolayers MnSTe and MnSeTe exhibit remarkable micromagnetic DMI values of 9.1 and 6.7 mJ/m$^2$, respectively. Another factor that affects the formation of skyrmions is the magnetic anisotropy energy $K_u$. The value of $|K_u|$ of MnSTe (MnSeTe) is $6.0\times10^6$ J/m$^3$ ($4.1\times10^6$ J/m$^3$), which is smaller than the values of $1.25*|\frac{D^2}{A}|$ of MnSTe (MnSeTe) $9.2\times10^6$ J/m$^3$ ($4.2\times10^6$ J/m$^3$). It means that the skyrmion states can be stabilized [41], indicating the possibility of creating skyrmions in Janus MnSTe and MnSeTe. Meanwhile, the $|K_u|$ in VSeTe ($4.4\times10^6$ J/m$^3$) is much larger than its values of $1.25*|\frac{D^2}{A}|$ ($1.3\times10^6$ J/m$^3$), which suggests the instability of the skyrmion in VSeTe [41].

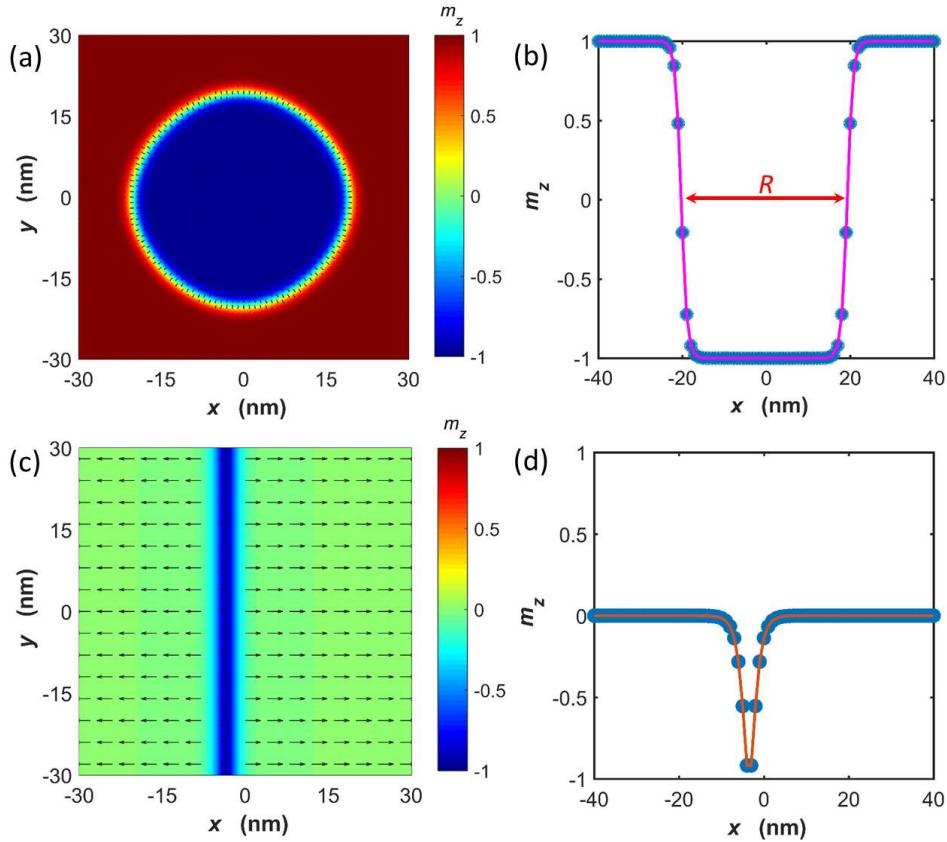

**FIG. 3.** Magnetization distribution for the relaxed states at zero magnetic field in Janus MnSTe (a)-(b) and VSeTe (c)-(d). (a) skyrmion profile in the xy plane, (b) out-of-plane magnetization component $m_z$ for the line profiles across the center of the skyrmions (y=0) for monolayer MnSTe. Magnetic domains profile of VSeTe in the (c) xy plane and (d) out-of-plane magnetization component $m_z$ for the line profiles across magnetic domains wall along

y=0. The color scale for m_z is inserted and arrows represent the in-plane component.

To further verify the aforementioned scenario, the micromagnetic simulations based on the magnetic parameters extracted from DFT are performed for MnSTe and VSeTe cases. The magnetization is initialized as a skyrmion-like state and the relaxed states at zero magnetic field are depicted [**Fig. 3**]. One can see that a stable Néel skyrmion with the magnetization of the center along the -z direction, opposite to that of ferromagnetic (FM) background at the boundary, is observed in monolayer MnSTe [**Fig. 3**]. It is straightforward to read the out-of-plane magnetization component $m_z(x)$ along y=0 [see **Fig. 3(b)**] and determine the diameter of the skyrmion $R$ [42]. The diameter of skyrmion without an external magnetic field in monolayer MnSTe is around 40 nm, which is smaller than that of Co/Pt [16]. The magnetic skyrmions are topologically stable with a quantized topological index $n$ being defined as $n = \frac{1}{4\pi} \int (\frac{\partial \boldsymbol{m}}{\partial x} \times \frac{\partial \boldsymbol{m}}{\partial y}) \cdot \boldsymbol{m} dx dy$, where $\boldsymbol{m}$ is the normalized magnetization vector, and x and y are the coordinates. The topological index for the skyrmions in this case of MnSTe is $n = -1$ at zero magnetic field. We also study VSeTe which has an in-plane magnetic anisotropy. It is found that the relaxed state for VSeTe is not a skyrmion but two in-plane magnetic domains with a domain wall as shown in **Fig. 3(d)** and **3(e)**. As discussed in the previous paragraph, Néel skyrmions cannot be stabilized in VSeTe because of its sizeable in-plane magnetic anisotropy, $K_u$ [see in **Table 1**].

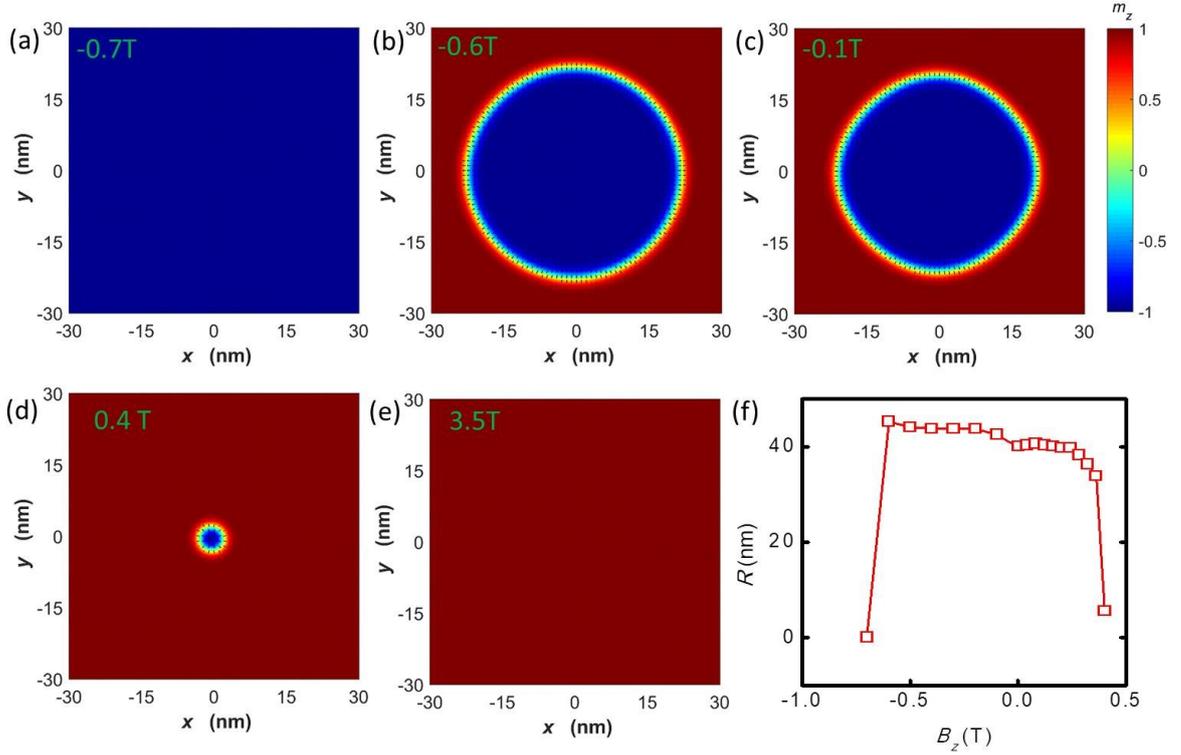

**FIG. 4.** The evolution of the size and shape of skyrmions as a function of the external magnetic field for MnSTe.

Furthermore, the impact of the external magnetic field onto the size and shape of skyrmions in monolayer MnSTe are also explored. The skyrmions at different external magnetic field strengths are mapped in **Fig. 4**. When the positive external magnetic field is applied, the rim of the skyrmion shrinks rapidly to its center with the increasing of the magnetic field. The size of skyrmions decreases with the increasing of the positive external magnetic field. The size of skyrmion reduces to 5.5 nm at 0.4 T and then collapses into a ferromagnetic state with all the spin along the positive direction at 3.5 T. The significant decreasing of spins component opposite to the magnetic field induces a qualitative change of the skyrmion shape under an external magnetic field. Moreover, the rim of the skyrmion expands slowly outward with the increasing of the magnetic field, when the negative external magnetic field is applied. The size of skyrmion increases with the increasing of the negative external magnetic field. Until the magnetic field is up to -0.7 T, the external magnetic field destroys the skyrmion state and finally turns into a collinear ferromagnetic state with all the spin along the negative direction. The stability of skyrmions against an external field is very high and the skyrmions can be stabilized under a large range of magnetic fields. The transition from the non-trivial skyrmion state to the trivial ferromagnetic state occurs in high magnetic fields with the topological index from n = −1 to n = 0. When the external magnetic field is applied, the subtle balance of all involved energies is broken and subsequently reconstructed. The Zeeman energy introduces a new external force, which makes the whole system tends to ferromagnetic state, while DMI competes with Zeeman energy and stabilizes skyrmions against collapse to a ferromagnetic state.

## IV. CONCLUSION

In summary, we have investigated the noncollinear magnetic properties of a family of Janus vdW magnets, MnSTe, MnSeTe, VSeTe, and MnSSe. All of them are ferromagnetic metals with Zener double-exchange interaction. Significant intrinsic DMI was found in MnSTe, MnSeTe, and VSeTe with the inversion asymmetry and strong SOC originated from the Te atoms. The micromagnetic simulations demonstrated that the Néel skyrmions could be stabilized in both monolayer MnSTe and MnSeTe in the absence of external magnetic fields. The diameters of skyrmions in monolayer MnSTe and MnSeTe are 40 and 41 nm, which could be shrunk to sub-10 nm by an external magnetic field. The Néel skyrmions exist under a large range of external magnetic field. In this work, we reveal the underlying physics of the existence of intrinsic magnetic skyrmions in atomic layered materials. Our research provides an alternative platform to discover/design intrinsic DMI and skyrmions in 2D vdW magnets for spin-orbitronic and memory devices. After submitting this paper, we became aware of two similar works [43,44]. By replacing half iodine atoms with Br or Cl, Xu *et al.* demonstrated helical cycloid and skyrmion phases in $CrI_3$ with the aid of magnetic fields through Monte Carlo simulations [43]. Liang *et al*, reported room-temperature Curie temperature, large DMI and skyrmion phases in Mn-based Janus dichalcogenides also via Monte Carlo simulations [44].

**ACKNOWLEDGEMENTS**

J.Y. and Y.Y. contributed equally to this work. This work is supported by the National Natural Science Foundation of China (NSFC91750112). J.Y. and L.S. acknowledge the support from Singapore MOE Tier 1 (Grant R-265-000-615-114 and R-265-000-651-114) and the computational resources provided by the Centre for Advanced 2D Materials (CA2DM) at the National University of Singapore.

**APPENDIX**

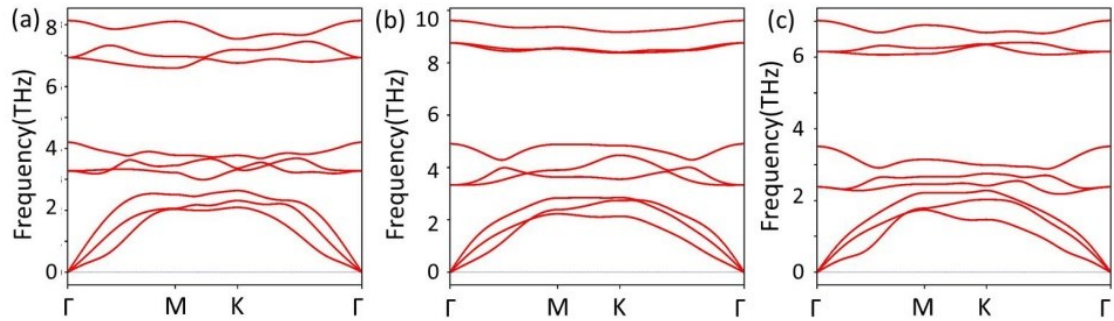

**FIG. 5**. The calculated phonon dispersion of monolayer Janus TMDs along the high symmetry path of the Brillioun zone: (a)MnSSe, (b) MnSTe, and (c) MnSeTe. The absence of imaginary frequency confirms that monolayer Janus TMDs are energetically stable.

The calculated phonon dispersions of monolayer Janus TMDs are plotted in the supporting information **Fig. 5**. The absence of imaginary frequency confirms that monolayer Janus TMDs are energetically stable.

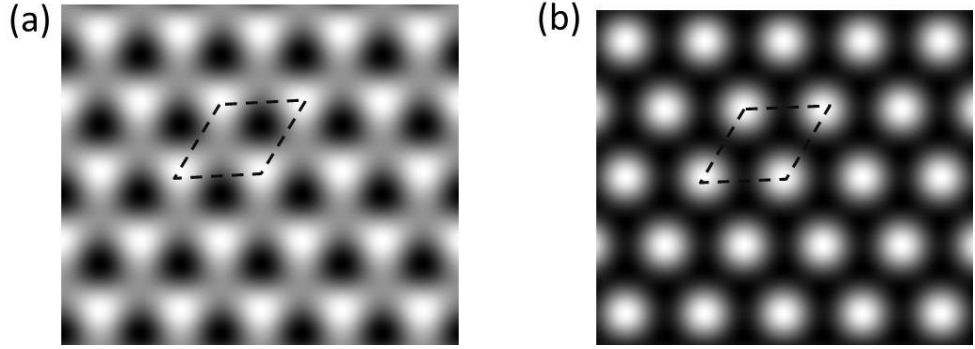

**FIG. 6**. The simulated STM image of (a) the pristine MnSe$_2$ and (b) Janus MnSeTe supercells under a bias voltage of 0.2 V.

**Fig. 6(a)** and **6(b)** are the simulated STM images of the pristine MnSe$_2$ and Janus MnSeTe supercell under a bias voltage of 0.2 V, respectively. It is clear that the intensity distribution of the two systems are different, indicating the electron states in the vicinity of the Fermi level are different for the pristine and Janus structure.

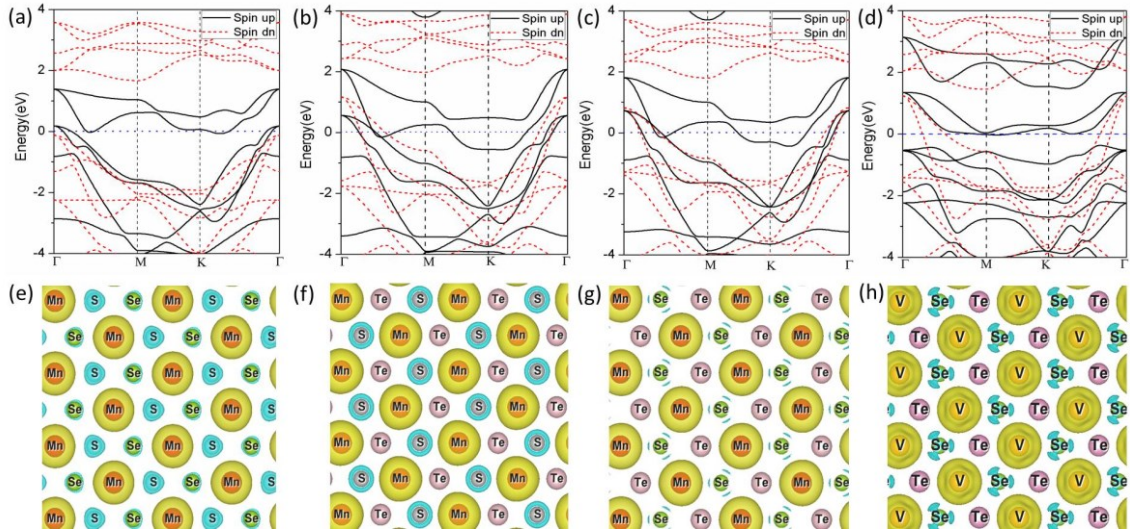

**FIG. 7.** The spin-polarized band structure of Mn-based Janus TMDs: (a) MnSSe, (b) MnSTe, (c) MnSeTe and (d) VSeTe. The spin up and spin down bands are indicated by black solid lines and red dashed lines, respectively. The Fermi level is set to zero and is represented by a blue dotted line. The spin-polarized charge density of Mn-based Janus TMDs: (e) MnSSe, (f) MnSTe, (g) MnSeTe and (h) VSeTe. The spin up and spin down charge are indicated by yellow and blue isosurface, respectively.

The spin-polarized band structure of Janus TMDs along the high symmetry line is plotted in **Fig. 7(a)-7(d)**. The energy band for the Janus TMDs passes through the Fermi level with large spin splitting. Hence, monolayer Janus TMDs present ferromagnetic metallic property, which is similar to that in the pristine 1T phase TMDs as shown in **Fig. 8**. Interestingly, in the case of

MnSSe, the spin-up electrons present a metallic nature while the spin-down electrons display a semiconducting property with a bandgap of 1.82 eV, which implies that monolayer MnSSe is a half-metal. Therefore, the electron transport for MnSSe is dominated by the spin-up states, and a 100% spin-filtering efficiency can be achieved near the Fermi level. The other there Janus structures are ordinary magnetic metals with both spin-up and spin-down bands passing through the Fermi level.

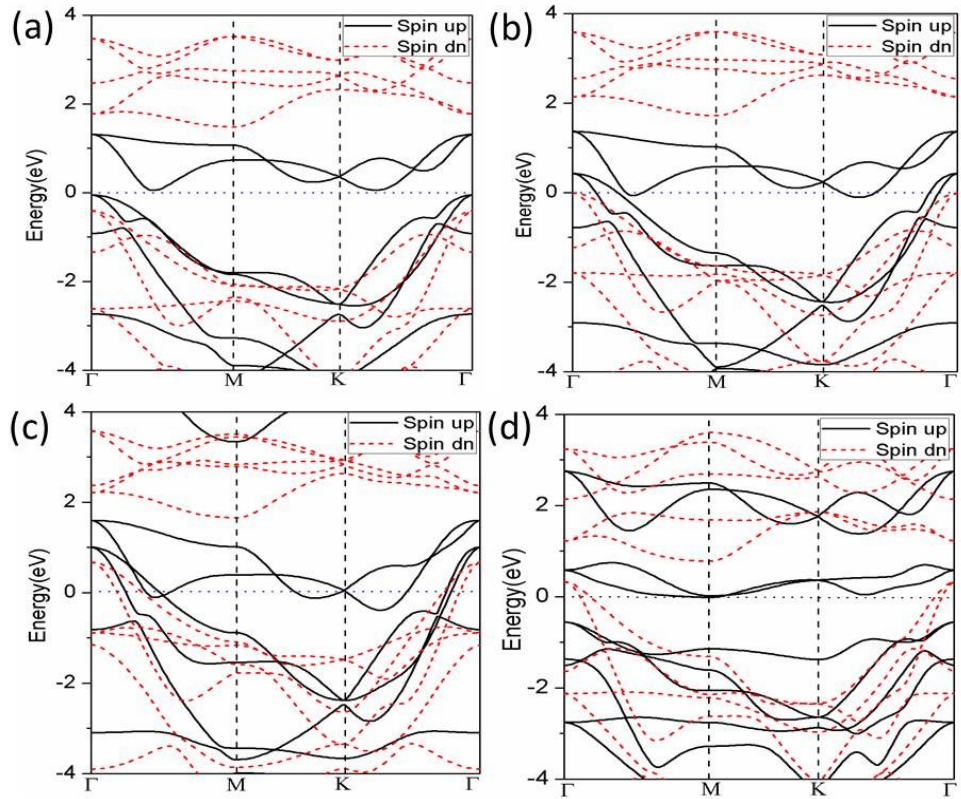

**FIG. 8.** The spin-polarized band structure of Mn-based pristine 1T TMDs: (a) $MnS_2$, (b) $MnSe_2$, (c) $MnTe_2$ and (d) $VSe_2$. The spin up and spin down bands are indicated by black solid lines and red dashed lines, respectively. The Fermi level is set to zero and represented by a blue dotted line. The Fermi level touches the energy band for all the pristine 1T TMDs with large spin splitting, indicating that monolayer pristine 1T TMDs are magnetic material.

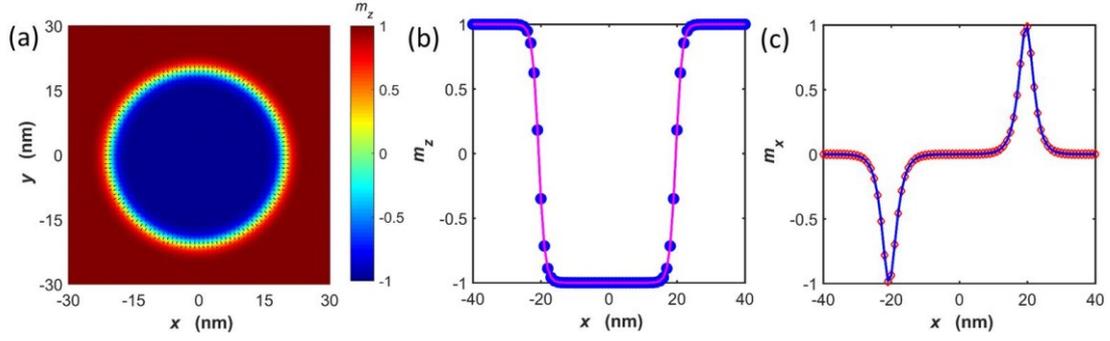

**FIG. 9.** (a) Magnetic domains profile in the xy plane, (b) out-of-plane magnetization component $m_z$ and (c) in-plane magnetization component $m_x$ for the line profiles across magnetic domains wall along y=0 in MnSeTe. The color scale for $m_z$ is inserted and arrows represent the in-plane component.

As plotted in **Fig. 9**, it is clear that a stable Néel skyrmion with $R$=41 nm is formed in monolayer MnSeTe without magnetic field.

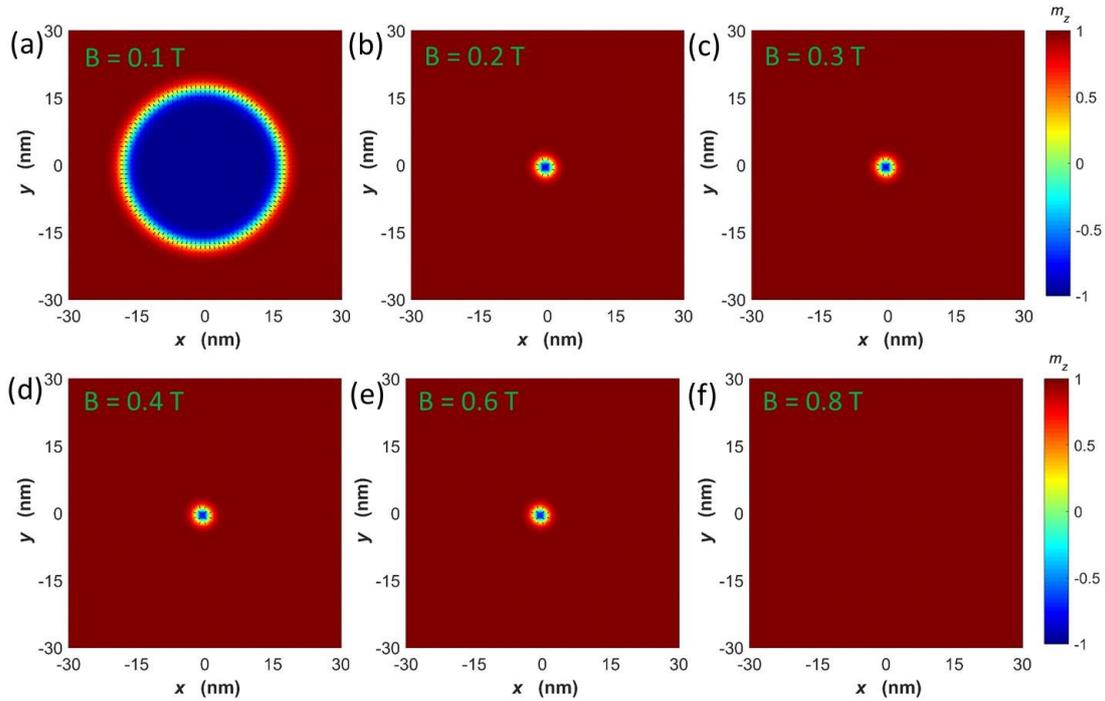

**FIG. 10.** The evolution of the size and shape of skyrmions as a function of positive external magnetic field for MnSeTe.

When the positive external magnetic field is applied as shown in **Fig. 10**, the rim of the skyrmion shrinks rapidly to its center with the increasing of the magnetic field. It is found that the size of skyrmions decreases with the increasing of the positive external magnetic field. The size of skyrmion reduces to around 3 nm at 0.2-0.6 T, and then collapses into a ferromagnetic state with all the spin along the positive direction at 0.8 T. On the other hand, when the negative external magnetic field is applied as mapped in **Fig. 11**, the rim of the skyrmion expands slowly outward with the increasing of the magnetic field. Until the magnetic field is up to -0.6 T, the skyrmion state transitions into a collinear ferromagnetic state.

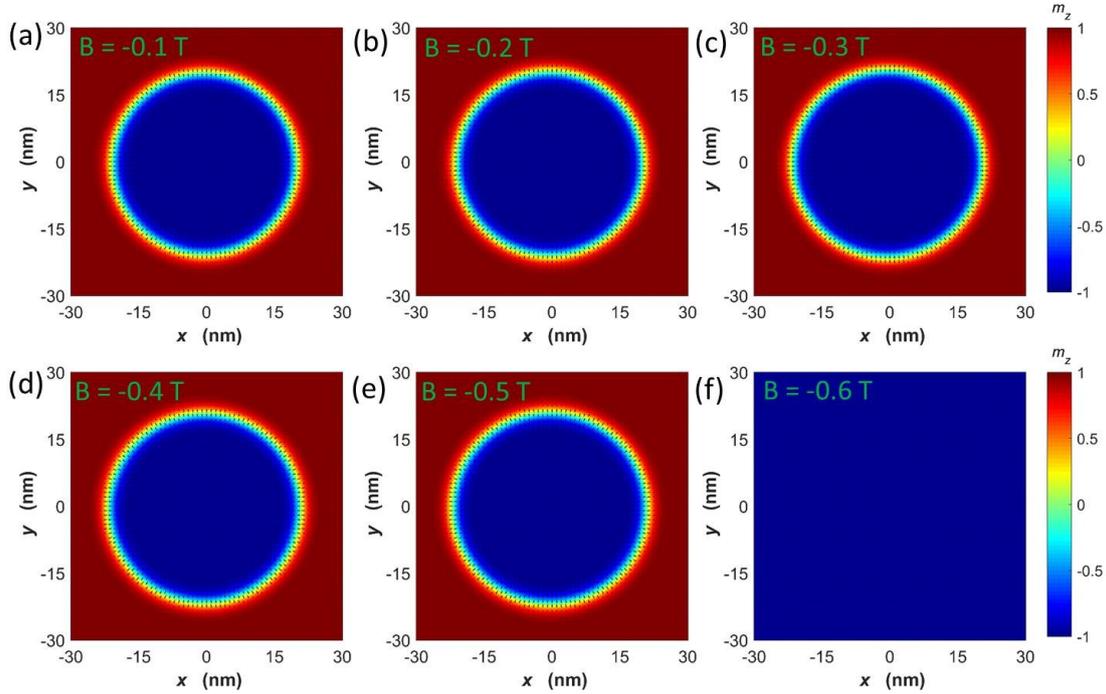

**FIG. 11.** The evolution of the size and shape of skyrmions as a function of negative external magnetic field for MnSeTe.

Email: yanxh@ujs.edu.cn

xuguiyin@mit.edu

shenlei@nus.edu.sg